\documentclass[12pt]{article}
\usepackage[utf8]{inputenc} 
\usepackage{geometry} 
\usepackage{graphicx}
\usepackage{physics}
\usepackage{tikz}
\usetikzlibrary{quantikz}
\usepackage{bm}
\usepackage{amssymb}
\usepackage{authblk}
\usepackage{hyperref}
\hypersetup{pdftex,colorlinks=true,allcolors=blue}
\usepackage{hypcap}
\usepackage{algorithm2e}
\usepackage{pgfplots}
\providecommand{\keywords}[1]{\textbf{\textit{Keywords:}} #1}

\begin{document}

\title{Quantum Algorithm to Estimate the Mean Value of a Function}

\author{Amanuel T. Getachew}
\affil{Department of Information Technology, Wolkite University, Ethiopia, amanuel.tamirat@wku.edu.et}

\maketitle

\begin{abstract}
This paper proposes a quantum circuit for computing the mean value from a given set of numbers or function evaluations. Suppose a Quantum Random Access Memory is given as a black-box function, which allows us to store and read the values of a set as quantum states. The proposed quantum algorithm estimate the mean value of the function by using superposition, interference, and entanglement phenomena, in $\mathcal{O}(\log{N})$ complexity or in $\mathcal{O}(1)$ query of the black-box.
\end{abstract}

\section{Introduction}
Quantum computers in principle, due to their wave-like properties, can solve hard computational problems much more efficiently than classical computers. Quantum computation is an entirely new way of computing. It concerns the investigation of computational power by the implications of quantum mechanics for information processing tasks. The art of developing algorithms for a potential quantum computer relies on some elementary gates to create a quantum state that has a relatively high amplitude for states that represent solutions for the given problem. A measurement in the computational basis then produces such a desired result with a relatively high probability.

One of the most exciting quantum algorithms discovered up to date is that quantum computers can find a unique key from the unstructured database in $\mathcal{O}(\sqrt{N})$ queries, which also known as Grover’s algorithm \cite{Grover1996}. Followed this, searching minimum or maximum value from a given unsorted list is explained by \cite{Durr1996b}, this algorithm achieves the same quadratic speedup as compared to the well-known classical version of it. 

Given $F(x)$, the mean estimation problem is to compute $\mu = \frac{1}{N} \sum_{x=0}^{N-1}{F(x)} $. When $F(x)$ is given as a black box (i.e. an oracle), the complexity of computing the mean can be measured by counting the number of queries made to this black box. This requires $N$ number of operations on a classical Turing machine. However, the nature of quantum mechanical systems allows quantum computers to easily solve this kind of statistical problems within a few numbers of steps \cite{Grover1996a}. For instance, \cite{Brassard2011} describes two algorithms to approximate the mean and median of the function based on amplitude estimation \cite{Brassard2000}, in $\mathcal{O}(\frac{1}{\epsilon} \log \log{\frac{1}{\epsilon}})$  or in $\Omega(\frac{1}{\epsilon})$ steps, whose complexity is measured based on the given presicion value, $\epsilon$ \cite{Nayak1999}.

This paper proposes a simple and efficient quantum circuit for computing the mean value from a given set of numbers or function evaluations, inspired by \cite{Schuld2017}. The proposed quantum algorithm differs from the previous works in that it is independent of Grover's algorithm. The rest of this paper is organized as follows. In the next section, the novel quantum mean-estimation algorithm is explained. Then, using the IBM QX, the results and the performance of the proposed quantum algorithm are discussed. Finally, we conclude the paper with a discussion and some perspectives on future extensions.

\section{Quantum Mean Estimation Circuit}
In this section, the proposed quantum mean estimation circuit is described.  The algorithm may assume a Quantum Random Access Memory (qRAM) as a black box function, to access the values of a set in parallel. Like a classical RAM, qRAM is also composed of two registers: address and data. But, unlike a classical RAM, data can be written or read in $\mathcal{O}(\log N)$ complexity \cite{Giovannetti2008a}. 

Let, $f(x)$ be a rescaled version of the original function $F(x)$, and assume it is given as a black box in which the black box accepts, $x$, one of the value from $\lbrack 0, 1, 2 ... N-1 \rbrack$, such that $x \in {\lbrace 0, 1 \rbrace}^{\log_{2} N}$; and produce an output $f(x)$ in the range -1 to 1 via a single qubit state $\ket{f(x)}$, where
\begin{equation} 
\ket{f(x)} = \sqrt{1-f(x)^2}\ket{0} + f(x)\ket{1}
\label{eq:fun}
\end{equation}
Without loss of generality, the above assumption should also work for a given set of numbers, where each of the $x^{th}$ values of the set are rescaled and stored in the qRAM as defined in Eq. \ref{eq:fun}.

The aim of the proposed quantum circuit is to estimate the average of the function, $\mu = \frac{1}{N} \sum_{x=0}^{N-1}{f(x)}$ with a little error overhead $\epsilon$. The algorithm encodes the final answer in the amplitude of a state through the interference pattern. Generally, without considering the complexity of the given function, the algorithm performs only $\mathcal{O}(\frac{1}{\epsilon})$ query of the function to return the required quantum state.

\begin{figure}[]
\begin{center}
\begin{quantikz}
\lstick{\textit{index register} $\ket{0}^{\otimes n}$} &[2mm] \gate{H}\qwbundle{n} & \gate[wires=2]{f(x)} & \gate{H} & \octrl{1} & \gate{H}& \qw \\
\lstick{\textit{data register} $\ket{0}$} & \qw  &  & \qw & \ctrl{1}& \qw & \qw \\
\lstick{\textit{mean register} $\ket{0}$} & \qw  & \qw & \qw & \targ{}& \qw & \qw & \meter{} \\
\end{quantikz}
\end{center}
\caption{Quantum circuit for mean estimation}
\label{fig:mean}
\end{figure}
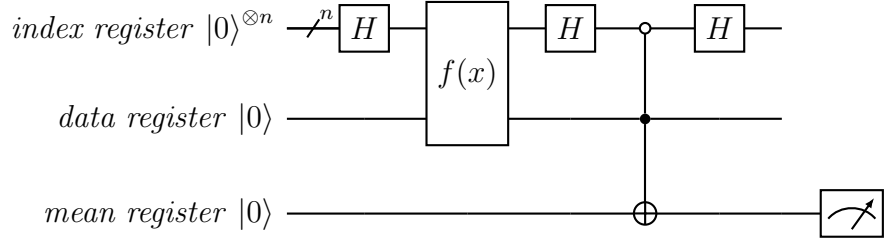

The general quantum circuit for the mean estimation algorithm shown in Fig \ref{fig:mean}. Three quantum registers are involved in the circuit, index register, data register, and mean register. The index register contains $n$ qubits where $n = \log{N}$  and it used to represent every $N$ evaluations. Both the data and mean registers have only one qubit in them. The data register is used to hold the value of the individual function evaluations, while the mean register is used to hold the output mean value.

The algorithm starts by setting up all of the qubits into the ground state and the system state at a time of zero can be exprssed as $ \ket{\psi_{0}} = \ket{0}^{\otimes n}\ket{0}\ket{0}$. In order to evaluate the function in superposition state, the Hadamard gate is applied on the index register. The operation changes the system state to 
\begin{equation} 
\ket{\psi_{1}} = \frac{1}{\sqrt{N}}\sum_{x=0}^{N-1}{\ket{x}}\ket{0}\ket{0}
\end{equation}
Then, the algotihm evaluate the function in parallel this makes the system state to include each values of the function
\begin{equation}
\ket{\psi_{2}} = \frac{1}{\sqrt{N}}\sum_{x=0}^{N-1}{\ket{x}} \ket{f(x)} \ket{0}
\end{equation}

The third step is creating interfence by applying the Hadamard gates on the index register this allows us to gain more information about the function. This operation brings the system state to 
\begin{equation} 
\ket{\psi_{3}} =\frac{1}{N}\sum_{x=0}^{N-1}{\ket{x}} \sum_{y=0}^{N-1}{(-1)^{x \cdot y} \ket{f(y)} } \ket{0}
\end{equation}
The data register value at this time is now changed. By factoring out the zero's component of the index register, we can rewrite the above equation as
\begin{equation} 
= \frac{1}{N}\left(\ket{0} \sum_{y=0}^{N-1}{\ket{f(y)}} + \sum_{x=1}^{N-1}{\ket{x}}\sum_{y=0}^{N-1}{(-1)^{x \cdot y}\ket{f(y)}}\right)\ket{0}
\end{equation}
When we see, the data register that entagled with the $\ket{0}^{\otimes n}$ component of the index register, it is simillar to the mean value of the function, $\ket{\mu}$. 

Next, using a general CNOT gate, the circuit extracts the value of data register that entagles with the zero's component of index register, and copies to the mean register. As the result, it creates entanglement between the two registers: data and mean. The system state at the time of four can expressed as 
\begin{equation} 
\ket{\psi_{4}} = \frac{1}{N}\left(\ket{0} \sum_{y=0}^{N-1}{\ket{f(y)}\ket{f(y)}} + \sum_{x=1}^{N-1}{\ket{x}}\sum_{y=0}^{N-1}{(-1)^{x \cdot y}\ket{f(y)}\ket{0}}\right)
\end{equation}

Finally, there is a last Hadamard transformation, that is applied to the index register qubits. This creates an interesting interference pattern especially on the mean register, see Eq. \ref{eq:last}. As we can see, the data register holds it's previous value, whereas, the mean register in the $\ket{1}$ component side has $\mu$ amplitude in every state of the system. 
\begin{equation}
\ket{\psi_{5}} = \frac{1}{\sqrt{N}} \sum_{x=0}^{N-1} \ket{x} \ket{f(x)}  \left( \frac{\hat{s} + \delta_{x}}{\sqrt{N}}  \ket{0} + \frac{s}{\sqrt{N}} \ket{1} \right)
\label{eq:last}
\end{equation}
where, $\hat{s} = \sum_{x=0}^{N-1}\sqrt{1-f(x)^2}$, $s = \sum_{x=0}^{N-1}f(x)$ and $\delta_{0} = N-1$ otherwise $\delta_{x} = - 1$.

This guarantees that upon the measurement of the mean register, we can estimate the mean value precisely. By counting the number of occurrences of being in the $\ket{1}$ state, dividing it by the number of shots yields to the square of the required result which is, $\mu^2$. In case if we want to know the sign of the mean value we can rerun the algorithm by slightly changing some of the input elements and measure the result again.

\section{Results}
To demonstrate the mean estimation algorithm, the proposed circuit is implemented in python, version 3.8 using the IBM Quantum Experience (IBM QX) module known as QISKit, version 0.23.2. The results of the two experiments are reported. Both experiments were simulated in an error-free environment (\textit{ibmq qasm simulator}) for the maximum number of 8192 shots.

In the first experiment, we considered the set of four numbers given as $[0.836, -0.549, 0.615, 0.053]$. For this small-scale example, a total of four qubits are required $i.e.$ two qubits for the index register and one qubit each for the data and mean registers. The quantum state representation for each input numbers encoded as
\begin{equation}
\ket{x_{0}} = 0.5487294415283365\ket{0} + 0.836\ket{1}
\end{equation}
\begin{equation}
\ket{x_{1}} = 0.8358223495456436\ket{0} -0.549\ket{1}
\end{equation}
\begin{equation}
\ket{x_{2}} = 0.7885271079677604\ket{0} + 0.615\ket{1}
\end{equation}
\begin{equation}
\ket{x_{3}} = 0.9985945123021657\ket{0} + 0.052999\ket{1}
\end{equation}
As expected the quantum circuits yield the probability of the mean register being in $\ket{1}$ around, $0.05743577075$. The mean value is then $\sqrt{0.0574357707} \to 0.239657611$ with the $\epsilon \to 0.000907611$.

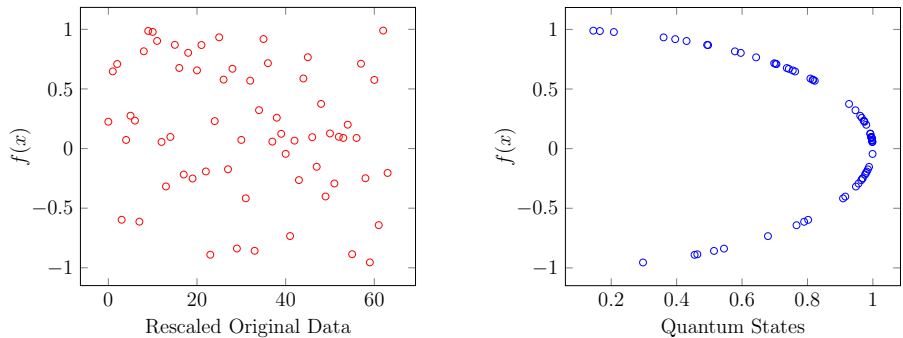
\begin{figure}[]
\begin{center}
\begin{tikzpicture}[scale=0.65] 
\begin{axis}[
scatter/classes={c={mark=o,draw=black, red}},
xlabel={Rescaled Original Data},
ylabel={$f(x)$}]
\addplot[scatter,only marks, scatter src=explicit symbolic]
coordinates {
(0, 0.22495386795899458) [c]
(1, 0.6472769072293065) [c]
(2, 0.709038451287777) [c]
(3, -0.5970796556344883) [c]
(4, 0.07235611518784857) [c]
(5, 0.27535441361958535) [c]
(6, 0.23405415872037216) [c]
(7, -0.6132848510586444) [c]
(8, 0.815838325977364) [c]
(9, 0.9862339892282083) [c]
(10, 0.9781021068377187) [c]
(11, 0.9026951005347351) [c]
(12, 0.05552000071686125) [c]
(13, -0.3178090994326095) [c]
(14, 0.09729156609423062) [c]
(15, 0.8698275786345279) [c]
(16, 0.6760137968713027) [c]
(17, -0.21776715526389767) [c]
(18, 0.8029093088538057) [c]
(19, -0.25104309810068315) [c]
(20, 0.6557036097200717) [c]
(21, 0.8682014099413626) [c]
(22, -0.1920455020733478) [c]
(23, -0.8905475758666307) [c]
(24, 0.23031742804316113) [c]
(25, 0.9329324998406741) [c]
(26, 0.5781468233293475) [c]
(27, -0.17386244011530716) [c]
(28, 0.669209363044652) [c]
(29, -0.8385876485402247) [c]
(30, 0.0724538958359131) [c]
(31, -0.4173503926375397) [c]
(32, 0.5690890246998732) [c]
(33, -0.857625412618098) [c]
(34, 0.32186644578510637) [c]
(35, 0.9183878964025863) [c]
(36, 0.7158760767584901) [c]
(37, 0.05801120581769359) [c]
(38, 0.258594374547069) [c]
(39, 0.12373176242755379) [c]
(40, -0.044414759114735414) [c]
(41, -0.7339100589130634) [c]
(42, 0.06636156341201288) [c]
(43, -0.26474588141090716) [c]
(44, 0.5877635730130948) [c]
(45, 0.7657913466078313) [c]
(46, 0.0954235152046492) [c]
(47, -0.1529304902589007) [c]
(48, 0.3747490078807687) [c]
(49, -0.40194124948724586) [c]
(50, 0.1268433647717387) [c]
(51, -0.29283200094197304) [c]
(52, 0.09849082135070958) [c]
(53, 0.08908499520159507) [c]
(54, 0.20018694857533098) [c]
(55, -0.8861796915682355) [c]
(56, 0.08889551420495267) [c]
(57, 0.7117261222084111) [c]
(58, -0.24912229271552627) [c]
(59, -0.9548132647575004) [c]
(60, 0.5756845545184449) [c]
(61, -0.6422214313117002) [c]
(62, 0.9893829417644175) [c]
(63, -0.2046396917481068) [c]
};
\end{axis}
\end{tikzpicture}
\qquad
\begin{tikzpicture}[scale=0.65] 
\begin{axis}[
scatter/classes={c={mark=o,draw=black, blue}},
xlabel={Quantum States},
ylabel={$f(x)$}
]
\addplot[scatter,only marks, scatter src=explicit symbolic]
coordinates {
(0.9743694152067209, 0.22495386795899436) [c]
(0.7622549477488905, 0.6472769072293065) [c]
(0.7051698196856064, 0.709038451287777) [c]
(0.8021819524443322, -0.5970796556344883) [c]
(0.9973788611129287, 0.0723561151878487) [c]
(0.9613427832465454, 0.2753544136195851) [c]
(0.9722235600856929, 0.23405415872037205) [c]
(0.7898618179542396, -0.6132848510586445) [c]
(0.5782800583337214, 0.815838325977364) [c]
(0.16535573316645072, 0.9862339892282083) [c]
(0.20812560774593766, 0.9781021068377187) [c]
(0.4302807867783366, 0.9026951005347352) [c]
(0.9984575752231036, 0.055520000716861945) [c]
(0.9481547217188943, -0.3178090994326094) [c]
(0.9952559224475542, 0.09729156609422997) [c]
(0.49335583856562826, 0.8698275786345279) [c]
(0.7368889647970345, 0.6760137968713027) [c]
(0.9760007510695212, -0.21776715526389756) [c]
(0.5961012009347942, 0.8029093088538057) [c]
(0.9679759102870333, -0.25104309810068304) [c]
(0.7550183946104014, 0.6557036097200717) [c]
(0.49621196254809297, 0.8682014099413626) [c]
(0.9813860224872758, -0.1920455020733477) [c]
(0.45489011323402884, -0.8905475758666307) [c]
(0.9731155544638999, 0.23031742804316124) [c]
(0.36005131681613195, 0.9329324998406741) [c]
(0.8159327488673219, 0.5781468233293475) [c]
(0.9847699487276971, -0.17386244011530747) [c]
(0.7430739050817026, 0.6692093630446521) [c]
(0.5447666984276631, -0.8385876485402248) [c]
(0.9973717626733768, 0.07245389583591318) [c]
(0.9087456463528678, -0.41735039263753965) [c]
(0.8222759159589603, 0.5690890246998732) [c]
(0.5142748794483717, -0.857625412618098) [c]
(0.9467850817781527, 0.32186644578510626) [c]
(0.39568127544935994, 0.9183878964025863) [c]
(0.6982273574738191, 0.7158760767584901) [c]
(0.9983159319572023, 0.058011205817692875) [c]
(0.9659859985800053, 0.25859437454706924) [c]
(0.9923157012597208, 0.12373176242755354) [c]
(0.9990131776772416, -0.04441475911473632) [c]
(0.6792466602245636, -0.7339100589130635) [c]
(0.9977956418533374, 0.0663615634120123) [c]
(0.964318214219747, -0.2647458814109074) [c]
(0.8090327448496015, 0.5877635730130947) [c]
(0.6430891178216005, 0.7657913466078313) [c]
(0.9954367648153187, 0.09542351520464872) [c]
(0.9882369478769615, -0.15293049025890068) [c]
(0.9271263026645181, 0.3747490078807686) [c]
(0.9156654585385602, -0.40194124948724574) [c]
(0.9919227594996415, 0.12684336477173885) [c]
(0.9561639081372609, -0.2928320009419729) [c]
(0.9951379593351178, 0.09849082135070941) [c]
(0.9960240276368496, 0.08908499520159527) [c]
(0.9797577178160414, 0.20018694857533106) [c]
(0.4633417251360242, -0.8861796915682355) [c]
(0.996040956765452, 0.08889551420495274) [c]
(0.7024570641442635, 0.7117261222084111) [c]
(0.9684720353588738, -0.24912229271552605) [c]
(0.29720637517241033, -0.9548132647575004) [c]
(0.8176718740967551, 0.5756845545184448) [c]
(0.7665191668601322, -0.6422214313117002) [c]
(0.14533201486798134, 0.9893829417644175) [c]
(0.9788373698226074, -0.20463969174810673) [c]
};
\end{axis}
\end{tikzpicture}
\end{center}
\caption{The set of 64 rescaled numbers are shown on the left side of the figure. The right side data points are the encoded version of the given data points in the form of quantum states.}
\label{fig:exp1}
\end{figure}

The second experiment data points are shown in Fig.\ref{fig:exp1}. The result shows that the probability of the mean register being in the $\ket{1}$ were equals to $0.023715415$. If we put this probability result in the square root, we get the mean of the given set with a little error overhead, $\epsilon$, which is $\sqrt{0.023715415} \to 0.153998101$. This is a good estimation when we see the classical mean value, which is $0.154619033$.

\section{Conclusion}
In this paper, we have used the powerful parallel computing abilities of a quantum computer to find the mean value from the given sequence of numbers. The proposed quantum mean estimator circuit runs in a 
$\mathcal{O}(\frac{1}{\epsilon}\log{N})$ complexity, where the complexity is measured by the time to prepare the qRAM. As compared to the classical version of it which takes $\mathcal{O}(N)$, the proposed quantum algorithm provides an exponential speedup. But, to recover the quantum result into the classical result efficiently we need to run the algorithm many times as needed.

The proposed quantum mean estimation algorithm is composed of three quantum registers and one classical register. The first quantum register is the index register and contains $\log{N}$ qubits, where $N$ is the size of the set, both second and third quantum registers contains one qubit in them. Similarly, the classical register contains one classical bit, used to store the value of the mean during measurement. The algorithm works as follows, first, we need to rescale the given set of numbers since each element of the set should be encoded in the amplitude of a qubit and store on the qRAM. This allows, the algorithm to access the qRAM or the black box in parallel. Then by using the Hadamard transformation interference pattern, it produces the mean of the set in $\mathcal{O}(1)$ query of the qRAM.

The algorithm could assist to solve many statistics related problems. For instance, by combination with a quantum minimum finding algorithm, it can be used to easily find a median of a set. The algorithm can also be improved to find the mean and median of the given set of vectors, which is a central problem in quantum nearest centroid classifier and k-medians algorithms.

\bibliographystyle{unsrt}
\bibliographystyle{splncs04}
\bibliography{MeanEstimation.bib}
\end{document}